\documentclass[reprint, twocolumn, superscriptaddress]{revtex4-1}
\usepackage{bm, amsmath, amsfonts, amssymb, braket}
\usepackage{subfigure}
\usepackage{color}
\usepackage{graphicx}
\usepackage{dcolumn} % Align table columns on decimal point

\newcommand{\ii}{\text{i}}

\usepackage{rotating} % for 90 degree rotated table and fig

\begin{document}
\title{Hermitian zero modes protected by nonnormality: Application of pseudospectra}
\author{Nobuyuki Okuma}
\email{okuma@hosi.phys.s.u-tokyo.ac.jp}
\author{Masatoshi Sato}
 %\altaffiliation{Department of Physics, University of Tokyo, Hongo 7-3-1, 113-0033, Japan}%Lines break automatically or can be forced with \\
\affiliation{%
 Yukawa Institute for Theoretical Physics, Kyoto University, Kyoto 606-8502, Japan
}%

\date{\today}
\begin{abstract}
Recently, it was established that there exists a direct relation between the non-Hermitian skin effects, -strong dependence of spectra on boundary conditions for non-Hermitian Hamiltonians-, and boundary zero modes for Hermitian topological insulators. On the other hand, in terms of the spectral theory, the skin effects can also be interpreted as instability of spectra for nonnormal (non-Hermitian) Hamiltonians. Applying the latter interpretation to the former relation, we develop a theory of zero modes with quantum anomaly for general Hermitian lattice systems. Our theory is applicable to a wide range of systems: Majorana chains, non-periodic lattices, and long-range hopping systems. We relate exact zero modes and quasi-zero modes of a Hermitian system to spectra and pseudospectra of a non-Hermitian system, respectively. These zero and quasi-zero modes of a Hermitian system are robust against a class of perturbations even if there is no topological protection. The robustness is measured by nonnormality of the corresponding non-Hermitian system. We also present explicit construction of such zero modes by using a graphical representation of lattice systems. Our theory reveals the presence of nonnormality-protected zero modes, as well as the usefulness of the nonnormality and pseudospectra as tools for topological and/or non-Hermitian physics.  
\end{abstract}
\maketitle
\section{Introduction}

Topology \cite{Kane-review, Zhang-review} and nonhermiticity \cite{Bender-98, Bender-02, Bender-review, Konotop-review, Christodoulides-review} are major concepts of recent condensed matter physics.
One of the most intriguing proprieties of topological physics is the bulk-boundary correspondence \cite{Hatsugai-93}, in which massless modes with quantum anomaly appear on the boundary of the insulating bulk corresponding to a nontrivial topological invariant.
The counterpart of the non-Hermitian physics is the non-Hermitian skin effect \cite{MartinezAlvarez-18,Torres-2019,YW-18-SSH,YSW-18-Chern,Kunst-18}, where the eigenspectrum of a non-Hermitian Hamiltonian strongly depends on the boundary condition owing to the non-Bloch nature \cite{MartinezAlvarez-18, YW-18-SSH, YSW-18-Chern, Kunst-18,Torres-2019, Gong-18, Lee-19, YM-19, Kunst-19, Borgnia-19,Zhang-19, OKSS-20}.

Recently, the correspondence between the winding number and skin effect has been pointed out \cite{Gong-18, Lee-19} and proved \cite{OKSS-20,Zhang-19}.
On the basis of this understanding, Ref. \cite{OKSS-20} has also proposed the notion of symmetry-protected skin effects together with higher-dimensional ones, and found the correspondence between them and boundary anomalous zero modes of topological insulators/superconductors. These findings give a unified understanding of topological and non-Hermitian physics. 

In this paper, we give a construction of Hermitian Hamiltonians with anomalous zero modes defined on general lattices such as long-range and nonperiodic systems by generalizing the relationship between topological zero modes and non-Hermitian skin effects. 
Noticing that the skin effects can also be interpreted as special cases of unstable spectra of nonnormal matrices $H$ (i.e., $[H,H^\dagger]\neq0$), we relate the general nonnormal spectra and pseudospectra \cite{Trefethen} to anomalous exact and quasi-zero modes of the corresponding Hermitian Hamiltonians, which do not necessarily have a topological characterization.
Instead of the topological protection, we regard the nonnormal pseudospectrum as the measure of the robustness against perturbations, which includes the notion of the topological protection as a special case.
Since the only requirement for our theory is the nonnormality, the crystalline periodicity is also an unnecessary assumption.
Our theory not only proposes the notion of the nonnormality-protected zero modes, but also adds nonnormal spectra and pseudospectra as tools for topological and/or non-Hermitian physics.
The correspondences between concepts of Hermitian anomalous zero modes and those of nonnormal matrices are summarized in Table \ref{table1}.

\begin{table*}[]
\caption{Correspondences between concepts of Hermitian anomalous zero modes and those of nonnormal matrices.}
\label{table1}
\begin{ruledtabular}
\begin{tabular}{ll}
Hermitian anomalous zero mode& Nonnormal matrix\\
\hline
$\bullet$ Topological zero modes& $\bullet$Non-Hermitian skin effects\\
Su-Schrieffer-Heeger model,& Hatano-Nelson model,\\
1D topological insulators (TIs) and superconductors (TSCs),&Symmetry-protected skin effects,\\
Defect zero modes in higher-dimensional TIs and TSCs, etc.&Higher-dimensional skin effects, etc.\\
$\bullet$Anomalous zero modes $\supset$ Topological zero modes&$\bullet$Nonnormal spectral instability $\supset$ Non-Hermitian skin effects\\
Hermitian system with anomalous zero modes,&Nonnormal network,\\
Exact zero modes,&Nonnormal spectra,\\
Quasi-zero modes.&Nonnormal pseudospectra.
\end{tabular}
\end{ruledtabular}
\end{table*}

This paper is organized as follows.
In Sec. \ref{skineffect}, we briefly review the non-Hermitian skin effects and their topological origin on the basis of the theory developed in Ref. \cite{OKSS-20}. There the relationship between the semi-infinite dense spectrum and the topological zero modes plays a crucial role in the skin effects. We also point out a subtle point about the exact and quasi-zero modes, which motivates us to introduce the pseudospectrum in the next section. 
In Sec. \ref{nonnormal}, we introduce and define several concepts of general square matrices including nonnormal matrices.
In particular, the $\epsilon$ pseudospectrum $\sigma_{\epsilon}(H)$, which is the set of spectra of $H$ with  $\epsilon$ perturbations, is introduced to characterize the unstable nature of nonnormal spectra.
We also discuss a possible application of the pseudospecta for the nonequilibrium dynamics.
In Sec. \ref{construction}, we present our main result. We give a construction of Hermitian Hamiltonians with anomalous zero modes from nonnormal matrices by noticing the similarity between semi-infinite dense spectra in Sec. \ref{skineffect} and nonnormal pseudospectra.
We discuss the robustness of the anomalous nature of those zero modes in terms of the behavior of the pseudospectra.
In Sec. \ref{example}, we construct the explicit examples of Hermitian zero modes from nonnormal networks.
The constructions are graphically performed by giving the weighted directed graphs.

\section{Topological origin of Non-Hermitian skin effects\label{skineffect}}

In this section, we briefly review the non-Hermitian skin effects and their topological theory developed in Ref. \cite{OKSS-20} and point out a new insight about the pseudospectrum.
In this reference, the authors and the collaborators showed that the mathematics of non-Hermitian skin effects is identical to that of the boundary zero modes of Hermitian topological insulators. This correspondence is the starting point for the theory developed in the later sections.

\subsection{Non-Hermitian skin effect and winding number}
As a typical example of non-Hermitian skin effect, we consider the Hatano-Nelson model.  
The Hatano-Nelson model~\cite{Hatano-Nelson-96, Hatano-Nelson-97} without disorder is given by 
\begin{equation}
H^{\rm (HN)} := \sum_i \left[ \left( t+g \right) c^{\dagger}_{i+1} c_i + \left( t-g \right)c^{\dagger}_{i} c_{i+1} \right]   
\end{equation}
with $t > 0$ and $g \in \mathbb{R}$, or in matrix representation, 
\begin{equation}
H^{\rm (HN)}:=
\begin{pmatrix}
0&t-g&0&\cdots\\
t+g&0&t-g&\cdots\\
0&t+g&0&\cdots\\
\vdots&\vdots&\vdots&\ddots
\end{pmatrix}.
\label{eq: Hatano-Nelson}
\end{equation}
In this section, we are interested only in the energy spectra of non-interacting Hamiltonians. The statistics of the creation and annihilation operators $(c,c^\dagger)$ is not important.
Since the following argument is based on the spectral theory of matrices, we adopt the matrix representation unless otherwise noted.
Under the periodic boundary condition (PBC), 
the energy spectrum of Eq. (\ref{eq: Hatano-Nelson}) is given by the dispersion relation calculated by the Fourier transform.
In this one-band model, the dispersion is equivalent to the Bloch Hamiltonian $H^{\rm (HN)} \left( k \right) = \left( t+g \right) e^{\ii k} + \left( t-g \right) e^{-\ii k}$, which forms an ellipse for $g\neq0$ in the complex-energy plane. 

Under the open boundary condition (OBC), the spectrum of Eq. (\ref{eq: Hatano-Nelson}) is drastically changed from the PBC one. To see this, we map $H^{\rm (HN)}_{\rm OBC}$ to the following Hermitian matrix $H'$ by a similarity transformation (imaginary gauge transformation in Refs. \cite{Hatano-Nelson-96,Hatano-Nelson-97}):
\begin{align}
H' &:= V_{r}^{-1} H^{\rm (HN)}_{\rm OBC} V_{r}\notag\\
&=
\begin{pmatrix}
0&\sqrt{|t^2-g^2|}&\cdots\\
\sqrt{|t^2-g^2|}&0&\cdots\\
\vdots&\vdots&\ddots
\end{pmatrix},
\end{align}
where $[V_r]_{i,j}=\delta_{ij}r^i$ with $r=\sqrt{|t+g|/|t-g|}$. Since the similarity transformation does not change the eigenspectra of finite matrices in general,
the energy spectrum of the Hatano-Nelson model with the OBC coincides with the real spectrum of the Hermitian matrix $H'$, in contrast to the case in the PBC.
This extreme sensitivity of the spectrum against the boundary conditions is called non-Hermitian skin effect because plane-wave-like eigenstates of the mapped Hamiltonian become boundary-localized modes in the original Hamiltonian. 

\begin{figure}[]
\begin{center}
　　　\includegraphics[width=6cm,angle=0,clip]{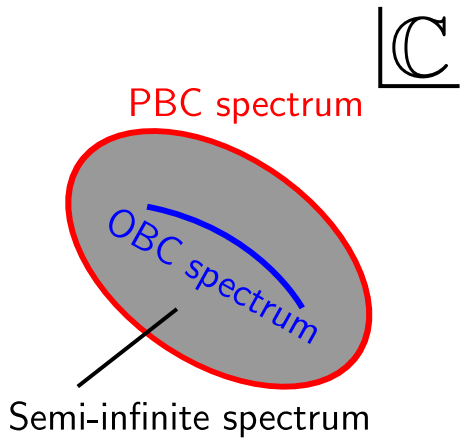}
　　　\caption{Complex spectra of a one-dimensional non-Hermitian Hamiltonian under the open (OBC), periodic (PBC), and semi-infinite boundary conditions. A semi-infinite dense spectrum indicates the presence of the non-Hermitian skin effect, where the OBC spectrum is far from the PBC one.}
　　　\label{fig1}
\end{center}
\end{figure}

The non-Hermitian skin effect can occur for one-dimensional non-Hermitian tight-binding models without any specific symmetry.
When a non-Hermitian Hamiltonian $H$ with finite-range hopping has the translation invariance in the bulk, the following statements hold in the infinite-volume limit~(Fig. \ref{fig1})~\cite{OKSS-20}:
\begin{itemize}
\item In the complex plane $\mathbb{C}$, the spectrum of $H$ under the semi-infinite boundary condition, where the boundary is only at the left-hand side of the system, is equal to the PBC spectral curve together with all the points $E\in \mathbb{C}$ enclosed by the PBC curve with the nonzero winding number $W(E)$.
\item For $W(E)<0$ [$W(E)>0$], the right (left) eigenstates of $H$ under the semi-infinite boundary condition exist and are localized at the boundary with exponential decay. 
\item The OBC spectrum is included in the semi-infinite spectrum.
\item The OBC curve has no winding.
\end{itemize}
The winding number $W(E)$ is defined as
\begin{equation}
W \left( E \right)
:= \oint_{C} \frac{d\beta}{2\pi \ii} \frac{d}{d\beta} \log \left( H \left( \beta \right) - E \right),\label{eq: winding number}
\end{equation} 
where $H(\beta)$ is the analytic continuation of the Bloch Hamiltonian $H(e^{\ii k})$ to the whole complex plane. 
Under the PBC, $\beta$ is nothing but the plane-wave solution $e^{\ii k}$, while under the OBC, $\beta$ is on the generalized Brillouin zone, which should be determined via the non-Bloch theory. \cite{YW-18-SSH, YSW-18-Chern, Kunst-18,YM-19, Kunst-19}. 
The first and second statements are variants of the Toeplitz index theorem in spectral theory \cite{Trefethen,Bottcher}.
Roughly speaking, the third one holds because the OBC spectrum can be regarded as the semi-infinite spectrum with the additional boundary condition at the right-hand side.
The fourth one can be shown by using the properties of the similarity transformation (imaginary gauge transformation) as well as the index theorem.
As a consequence of these statements, the following theorem holds \cite{OKSS-20}.
\\

{\bf Theorem}~~The OBC curve has no winding ($W=0$). Thus, if the PBC curve has winding ($W\neq0$), the non-Hermitian skin effect, where the OBC spectrum is far from the PBC one, inevitably occurs.\\

\subsection{Semi-infinite boundary states and Hermitian topological zero modes}
As noted above, the non-Hermitian skin effect is explained in terms of the nontrivial winding number.
Actually, this number is nothing but the topological invariant of a one-dimensional Hermitian topological insulator with the chiral symmetry, as discussed below. As a consequence of the bulk-boundary correspondence in Hermitian topological insulators, the non-Hermitian localized modes in the semi-infinite system can be related to the Hermitian topological zero modes.

To see this, let us consider the non-Hermitian Hamiltonian $H$ defined on semi-infinite system and the following doubled Hermitian Hamiltonian with a reference point $E\in\mathbb{C}$ \cite{Feinberg-97}:
\begin{equation}
\tilde{H}_E := \left( \begin{array}{@{\,}cc@{\,}} 
	0 & H-E \\
	H^{\dag} - E^{*} & 0 \\ 
	\end{array} \right).
		\label{eq: extended - Hermitian}
\end{equation}
By construction, $\tilde{H}_E$ respects additional chiral symmetry: 
\begin{align}
    \Gamma \tilde{H}_E \Gamma^{-1}=-\tilde{H}_E
    ~~{\rm with}~~\Gamma=
    \begin{pmatrix}
    1&0\\
    0&-1
    \end{pmatrix}.
\end{align}
Thus, the non-Hermitian physics without symmetry (class A in Altland-Zirnbauer classification \cite{Altland}) can be related to the Hermitian physics with the chiral symmetry (class AIII) via Eq. (\ref{eq: extended - Hermitian}).
An important by-product of this construction is the discovery that the aforementioned index theorem in the spectral theory for the class-A non-Hermitian Hamiltonians is equivalent to the bulk-boundary correspondence in the one-dimensional class-AIII Hermitian topological insulator because the winding number in Eq. (\ref{eq: winding number}) also gives the explicit expression of the $\mathbb{Z}$ topological invariant of $\tilde{H}_E$. 
In other words, the non-Hermitian localized boundary modes with the reference energy $E\in\mathbb{C}$ correspond to the one-dimensional class-AIII topological zero modes of $\tilde{H}_E$ as discussed below.

According to the conventional bulk bulk-boundary correspondence, $\tilde{H}$ possesses topologically protected zero modes localized at the boundary~\cite{Kane-review, Zhang-review, Schnyder-Ryu-review}.
For $W \left( E \right) < 0$, there appear boundary modes with negative chirality:
\begin{align}
    \tilde{H}_E
    \begin{pmatrix}
    0\\
    |E\rangle
\end{pmatrix}&=0,\\
\Gamma
    \begin{pmatrix}
    0\\
    |E\rangle
    \end{pmatrix}&=-
    \begin{pmatrix}
    0\\
    |E\rangle
    \end{pmatrix},
\end{align}
which implies that $\ket{E}$ is a right eigenstate of non-Hermitian $H$ (i.e., $H \ket{E} = E \ket{E}$) localized at the boundary. For $W \left( E \right) > 0$, on the other hand, the boundary modes have positive chirality:
\begin{align}
    \tilde{H}_E
    \begin{pmatrix}
    |E\rangle\\
    0
\end{pmatrix}&=0,\\
\Gamma
    \begin{pmatrix}
    |E\rangle\\
    0
    \end{pmatrix}&=
    \begin{pmatrix}
    |E\rangle\\
    0
    \end{pmatrix},
\end{align}
which in turn implies that $\ket{E}$ is a right eigenstate of $H^{\dag}$, i.e., a left eigenstate of $H$ (i.e., $\bra{E} H = \bra{E}E$)~\cite{Brody-14}. 
In both cases, the semi-infinite boundary modes of non-Hermitian $H$ are constructed from the topological zero modes of Hermitian $\tilde{H}$.
In terms of this correspondence, $\tilde{H}$ of the Hatano-Nelson model is nothing but the Su-Schrieffer-Heeger model~\cite{SSH-79}.

The above discussion is valid for arbitrary $E \in \mathbb{C}$ satisfying $W \left( E \right) \neq 0$. Thus, in semi-infinite systems, an infinite number of boundary modes emerge as a result of the nontrivial winding number $W \left( E \right) \neq 0$, which can be related to the boundary zero modes of a topological insulator.

\subsection{Skin effect and Hermitian topological zero modes: A subtlety about exact and quasi-zero modes}
According to the above discussion, there is a direct correspondence between non-Hermitian semi-infinite boundary modes and Hermitian topological zero modes.
Basically, this relation can be generalized to the full OBC case, where there is an additional boundary condition at the right-hand side of the one-dimensional system.
In this case, however, there arises a subtle point.
In physics, one believes that the class-AIII bulk-boundary correspondence predicts the presence of topological zero modes for any $E$ satisfying $W(E)\neq0$ even for the OBC, while the non-Hermitian skin modes are present only for a particular set of $E$. The origin of this mismatch is that the former includes the quasi-zero modes that become exact zero modes only in the infinite-volume limit. 
By construction, the direct correspondence between non-Hermitian boundary modes and Hermitian topological zero modes holds only for the exact zero mode.

This subtle point becomes essential when we consider the variants of the non-Hermitian skin effect in higher dimensions. In Ref. \cite{OKSS-20}, the authors and the collaborators generalized the non-Hermitian skin effect to other symmetry classes and dimensions, which we call symmetry-protected skin effect and higher-dimensional skin effect, respectively. We also constructed concrete examples of the one- and two-dimensional time-reversal-symmetric skin effects. In the two-dimensional example, the non-Hermitian skin effect does not occur under the full OBC, while it occurs for the case with the PBC in one direction and the OBC in the other direction. This comes from the fact that in a two-dimensional class-DIII superconductor, which corresponds to $\tilde{H}_E$, ``exact" topological zero modes are absent in the former boundary condition but they are present in the latter boundary condition. In the next subsection, we discuss this point for details.

Thus far, we have discussed the correspondence between ``exact" zero modes and the non-Hermitian skin modes.
The next question is, what the counterpart of the quasi-zero mode is.
The answer is the pseudospectrum, which is a generalization of the spectrum as defined in the next section. Roughly speaking, we can find the correspondence between a pseudo-eigenmode of the non-Hermitian system, which is an almost eigenstate with a slight deviation, and the topological quasi-zero mode. 
In this sense, we can still find the skin modes even though the doubled Hermitian Hamiltonian has no exact topological zero mode as in the case of the above example under the full OBC.

\subsection{Remarks about generalized skin effects from various topological zero modes}
As mentioned above, the notion of the non-Hermitian skin effect can be generalized to other symmetry classes and dimensions \cite{OKSS-20,Okuma-19}. Reference \cite{OKSS-20} pointed out that subsets of non-Hermitian classifications \cite{Gong-18,KSUS-19} describe such generalized skin effects.
The condition for this correspondence is that the doubled Hermitian Hamiltonian has the topologically-protected exact zero modes that lead to the semi-infinite dense spectrum in the original non-Hermitian Hamiltonian. In the case of the [$D(>1)$]-dimensional topological insulators, it is known that the topological zero modes under the full OBC cannot be exactly zero and have finite energies proportional to the inverse of the system size. Under the presence of the topological defect such as the $\pi$ flux in two dimensions, there can exist exact boundary and defect zero modes ~\cite{Qi-09}. As a result, the corresponding non-Hermitian Hamiltonian has the skin modes at the boundary and defect. 
For example, the two-dimensional time-reversal-symmetric skin effect under the full OBC can be induced by the $\pi$-flux insertion.

\section{Spectrum and Pseudospectrum of nonnormal matrix\label{nonnormal}}
The main purpose of this paper is to give a new construction of Hermitian anomalous zero modes by generalizing the relationship between topological zero modes and non-Hermitian skin effects.
For this purpose, we notice the fact that the non-Hermitian skin effects can also be understood as special examples of unstable spectra of nonnormal matrices $H$ (i.e., $[H,H^\dagger]\neq0$). On the basis of Ref. \cite{Trefethen}, we here briefly introduce the several concepts in the spectral theory including nonnormal pseudospectra, which will play a similar role as the semi-infinite spectra in the next section.
As an application of pseudospectra, we also discuss the nonequilibrium dynamics governed by the nonnormality.
\subsection{Nonnormal matrix and its measure}
In usual quantum mechanics, one focuses on Hamiltonians that are described by Hermitian matrices, which are specific examples of normal matrices $H$, i.e., $[H,H^\dagger]=0$.
The necessary and sufficient condition for $H$ to be normal is that $H$ can be written in the following form:
\begin{align}
    H=UDU^{-1}=UDU^{\dagger},
\end{align}
where $U$ is a unitary matrix, and $D$ is a diagonal matrix whose elements are eigenvalues of $H$.
Since $H$ can be diagonalized by a unitary matrix, there is no need to distinguish between right and left eigenvectors defined as
\begin{align}
    H|r,i\rangle&=E_i|r,i\rangle,\notag\\
    \langle l,i|H=\langle l,i|E_i&\Leftrightarrow H^{\dagger}|l,i\rangle=E^*_i|l,i\rangle,\label{leftrighteig}
\end{align}
where $i$ denotes the eigenvalue index.

Except for simple cases such as a system with a constant dissipation term,
non-Hermitian systems are described by nonnormal matrices $H$, i.e., $[H,H^\dagger]\neq0$.
Thus, the Hamiltonians are no longer diagonalized by unitary matrices and can be non-diagonalizable in extreme cases.
As a consequence, there exist pairs of left and right eigenvectors such that
\begin{align}
   |\langle l,i|r,i\rangle|<1,\label{biortho}
\end{align}
where the normalization is defined as $\langle r,i|r,i\rangle=\langle l,i|l,i\rangle=1$.
The presence of such eigenvectors is a unique property of nonnormal matrices. We use this property to define the Hermitian anomalous zero modes in the next section.

For a moment, let us assume that the nonnormal $H$ is diagonalizable, i.e., $H=PDP^{-1}$ with
\begin{align}
    P&:=\left(|r,1\rangle,|r,2\rangle,\cdots\right),\label{normp}\\
    P^{-1}&=
    \begin{pmatrix}
        \langle l,1|/\langle l,1|r,1\rangle\\
        \langle l,2|/\langle l,2|r,2\rangle\\
        \vdots
    \end{pmatrix}.
\end{align}
The nonnormality is often measured by $\|P^{-1}\|$, where $\|\cdot\|$ is the matrix norm. We adopt the 2-norm as the matrix norm:
\begin{align}
    \|A\|_2:=\max_{\bm{x}}\frac{\|A\bm{x}\|_2}{\|\bm{x}\|_2},
\end{align}
where $\|\bm{x}\|_2:=(\sum_i|x|_i^2)^{1/2}$ is the vector 2-norm.
By using this definition, the nonnormality is calculated in terms of the quantity in Eq. ($\ref{biortho}$):
\begin{align}
    \|P^{-1}\|=\frac{1}{\min_i |\langle l,i|r,i\rangle|}.
\end{align}
When the Hamiltonian is normal, $P=U$ and thus this quantity is unity.
An interesting nontrivial example is again the Hatano-Nelson model introduced in the previous section.
This model under the PBC is described by a normal matrix, and thus this quantity is unity.
Under the OBC, on the other hand, the system is nonnormal, and this quantity becomes infinity in the infinite-volume limit since it is the inverse of the overlap integral between a right eigenstate localized at one side and a left eigenstate localized at the other side. 

Note that we assume $\|P\|\sim 1$ by using the normalized eigenvectors [see also Eq. (\ref{normp})].
More generally, it is convenient to use the condition number \footnote{Note that $\kappa$ is not uniquely determined for given $H$ since $P$ is not unique. See Ref. \cite{Trefethen} for details.}: 
\begin{align}
    \kappa(P):=\|P\|\|P^{-1}\|.\label{conditionnum}
\end{align}
Since the 2-norm of a matrix is its largest singular value and the norm of the inverse is the inverse of the smallest singular value, the condition number is calculated as
\begin{align}
    \kappa(P)=s_{\max}(P)/s_{\min}(P)\geq1.
\end{align}
This also becomes unity only when $H$ is normal.
When $H$ is not diagonalizable, $\kappa$ is set to be infinite as a convention.

\subsection{Spectral theory of nonnormal matrices}
\begin{figure}[]
\begin{center}
　　　\includegraphics[width=8cm,angle=0,clip]{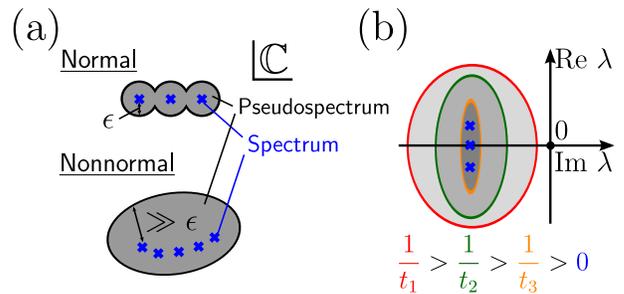}
　　　\caption{(a) Spectra and pseudospectra of normal and nonnormal matrices. (b) Lindblad spectrum and pseudospectra with $\epsilon=1/t_1,1/t_2,1/t_3$.}
　　　\label{fig2}
\end{center}
\end{figure}

The spectra of nonnormal matrices tend to be unstable against small perturbations.
For instance, the OBC spectrum of the Hatano-Nelson model, which is described by a nonnormal matrix as mentioned above, gets close to the PBC one under a nonlocal perturbation that connects two boundaries. Such an instability is well understood in terms of $\epsilon$ pseudospectra of the matrices. In this subsection, we introduce the notion of pseudospectrum and summarize some basic properties \cite{Trefethen}.

There are three identical definitions of $\epsilon$ pseudospectrum $\sigma_\epsilon(H)$ of a matrix $H\in \mathbb{C}^{N\times N}$ for arbitrary $\epsilon>0$:
\begin{itemize}
\item The set of $z\in\mathbb{C}$ such that $\|(z-H)^{-1}\|>\epsilon^{-1}$.
\item The set of $z\in\mathbb{C}$ such that $z\in\sigma(H+\eta)$ for some $\eta\in\mathbb{C}^{N\times N}$ with $\|\eta\|<\epsilon$.
\item The set of $z\in\mathbb{C}$ such that $\|(z-H)\bm{v}\|<\epsilon$ for some $\bm{v}\in\mathbb{C}^{N}$.
\end{itemize}
Here $\sigma(\cdot)$ is the spectrum of the matrix.
By definition, the pseudospectrum describes behaviors of spectra under perturbations.
In the case of Hermitian matrices, any perturbations to them do  no change their spectra so much.
Actually, for general normal matrices, the pseudospectrum is given by the $\epsilon$ neighborhood of the spectrum [Fig. \ref{fig2}(a)] \cite{Trefethen}:
\begin{align}
    \sigma_\epsilon(H)=\sigma(H)+\Delta_\epsilon:=\{z~|~{\rm dist}(z,\sigma(H))<\epsilon\},
\end{align}
where dist($\cdot,\cdot$) denotes the distance between two points in the complex plane.  
In the case of nonnormal matrices, on the other hand, the pseudospectrum is larger than the $\epsilon$ neighborhood of the spectrum [Fig. \ref{fig2}(a)]:
\begin{align}
     \sigma_\epsilon(H)\supset\sigma(H)+\Delta_\epsilon.\label{epsilonpseudo}
\end{align}
Equation (\ref{epsilonpseudo}) means that small perturbations to nonnormal matrices drastically change the spectrum.
In general, the upper bound of the pseudospectrum of a diagonalizable matrix is given in terms of the condition number (\ref{conditionnum}) \cite{Trefethen}:
\begin{align}
     \sigma(H)+\Delta_\epsilon\subseteq\sigma_\epsilon(H)\subseteq\sigma(H)+\Delta_{\kappa(P)\epsilon}.
\end{align}
In other words, the nonnormality measures the instability of the spectrum against small perturbations.
\subsection{Semi-infinite spectra versus pseudospectra}
As we discussed in the previous section, the spectrum of a one-dimensional short-range semi-infinite tight-binding model without symmetry (class A) is given by the corresponding PBC spectrum together with all the points enclosed by the PBC curve with nonzero winding number.
This ``dense" nature of the semi-infinite spectrum looks like the behavior of the nonnormal pseudospectra.
In fact, the following relation between the semi-infinite spectrum and the corresponding OBC pseudospectrum holds:
\begin{align}
    \sigma(H_{\rm SIBC})=\lim_{\epsilon\rightarrow0}\lim_{N\rightarrow\infty}\sigma_\epsilon(H_{\rm OBC}).\label{semi-pse}
\end{align}
This relation is compatible with the fact that nonlocal perturbations that connect two ends of a one-dimensional system can drastically change the OBC spectrum \cite{Xiong-2018}.
Note that these two limits do not commute, and the opposite order of limit corresponds to the OBC spectrum in the infinite-volume limit.

\subsection{Possible application for nonequilibrium phenomena}
There are lots of applications of pseudospectra.
For example in fluid mechanics, the stability of some fluid flow can be well described not by the spectrum but by the pseudospectrum of the linearized differential equation, which indicates that the nonnormality is important as well as the nonlinearity \cite{Trefethen}. In network science, which treats the complex systems in physics, biology, and sociology, the transient dynamics is governed by the pseudospectra of nonnormal networks \cite{Asllani-2018}.
In particular, Gong $et$ $al$. discussed the relationship between the boundary quench and Lieb-Robinson bound in the Hatano-Nelson model and claimed that quasi-edge modes behave like eigenstates up to some time scale \cite{Gong-18}. 

Although it is not directly related to the main subject, we here discuss a possible application of pseudospectra for the Lindblad superoperators \cite{Prosen-2008,Prosen-2010} [Fig. \ref{fig2}(b)], which describe nonequilibrium quantum phenomena in certain conditions.
The imaginary part of spectra \footnote{Note that the convention about the factor i changes the real and imaginary axes.} contains the information that determines the relaxation processes against the nonequilibrium steady state.
The eigenvalue with the largest imaginary part (the spectral abscissa in mathematics) is known to govern the long-time behavior ($t\rightarrow\infty$), while the largest eigenvalue of the anti-Hermitian part of the matrix (the numerical abscissa) is expected to describe the short-time quench dynamics ($t\rightarrow0$).
We expect that pseudospectra describe the transient dynamics ($0<t<\infty$) of the relaxation process.
For each $\epsilon>0$, the pseudoeigenvalue with the largest imaginary part, which corresponds to the pseudospectral abscissa [$\alpha_\epsilon(A)$], characterizes the time-scale of the relaxation if $\epsilon$ is sufficiently small. Roughly speaking, the competition between $1/\epsilon$ and $1/|\alpha_\epsilon(A)|$ determines the transient region of the dynamics. Thus, the strong nonnormality that makes the pseudospectral abscissa for small $\epsilon$ far from the spectral abscissa affects the transient but relatively long-time dynamics, for example in the systems with non-Hermitian skin effects.
It would be interesting to compare it with the recent work about the true long-time Lindblad dynamics of the non-Hermitian skin effect \cite{Song-19-Lindblad}.

The notion of pseudosepctrum is also related to the retarded and advanced Green functions:
\begin{align}
    G^{R(A)}(\omega)=\frac{1}{\omega-H\pm\Sigma},
\end{align}
where $\omega\in\mathbb{R}$ is the frequency, and $H$ and $\Sigma$ are the non-interacting Hermitian Hamiltonian and self-energy, respectively.
Apparently, the form of the Green functions is the same as the resolvent in the first definition of the pseudospectrum.
Since Green functions appear in the expressions of transport quantities, we expect that there exist transport phenomena whose origin is the nonnormality. 
%In particular, the form of the spectral function and the density of states respect the behavior of the pseudospectra on the real axis.

\section{Construction of Hermitian zero modes from nonnormal matrices\label{construction}}
One of the most intriguing properties of the topological insulators is the anomalous boundary zero mode, which is related to the quantum anomaly. In this section, we construct anomalous zero modes from general nonnormal matrices, which do not necessarily have the topological characterization. Although such zero modes are not always protected by bulk invariants, the anomalous nature is still robust against certain symmetry-preserving perturbations. 
We regard the nonnormal pseudospectral behavior as the measure of the robustness against certain perturbations.
Since our theory includes the notion of the topological protection as a special case, we use the correspondence between the Hatano-Nelson model and Su-Schrieffer-Heeger model as a concrete example for the comprehensive understanding of several concepts.
We first consider the class-AIII case as the simplest example. We then generalize the theory to the class-BDI case, where the anomalous zero modes correspond to the Majorana fermions, and discuss other symmetry classes.

\subsection{Class-AIII exact and quasi-zero modes from nonnormality}
\begin{figure}[]
\begin{center}
　　　\includegraphics[width=6cm,angle=0,clip]{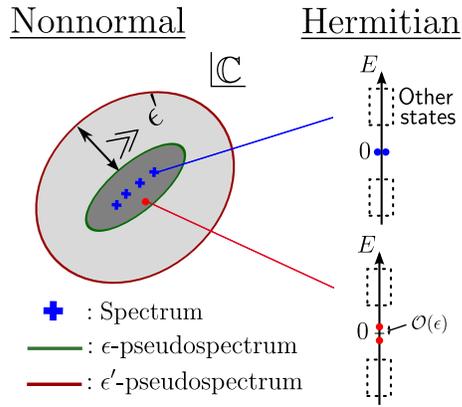}
　　　\caption{Correspondence between nonnormal concepts and Hermitian zero modes. Exact and quasi-zero modes can be constructed from nonnormal spectrum and pseudospectrum.}
　　　\label{fig3}
\end{center}
\end{figure}

As in the case of Eq. (\ref{eq: extended - Hermitian}), we can trivially construct a class-AIII Hermitian Hamiltonian with exact zero modes from a general square matrix $H$ that describes arbitrary class-A lattice systems with arbitrary spatial dimensions, and its eigenvalue $E$: 
\begin{align}
&\tilde{H}_E := 
\begin{pmatrix}
	0 & H-E \\
	H^{\dag} - E^{*} & 0 \\ 
\end{pmatrix},\notag\\
&\tilde{H}_E
\begin{pmatrix}
    0\\
    |r,E\rangle
\end{pmatrix}=0,~
\tilde{H}_E
\begin{pmatrix}
    |l,E\rangle\\
    0
\end{pmatrix}=0,
\end{align}
where $l$ and $r$ again denote the left and right eigenvectors of $H$ [see Eq. (\ref{leftrighteig})].
The first and second zero modes have the negative and positive chirality, as already mentioned.
These zero modes, however, are not always extracted as the anomalous zero modes isolated in real space.
For example, if $H$ is a normal matrix, $|r,E\rangle$ is identical to $|l,E\rangle$, and the two zero modes are located at the same position in real space, while two anomalous boundary zero modes found in the Su-Schrieffer-Heeger model are isolated from each other in real space. 

The conditions for these zero modes to be anomalous in real space are given as follows:
\begin{itemize}
\item $H$ is a nonnormal matrix, or equivalently, the $\epsilon$ pseudospectrum of $H$ is larger than the $\epsilon$ neighborhood of the spectrum of $H$.
\end{itemize}
This condition enables us to choose $|r,E\rangle$ and $|l,E\rangle$  such that $|\langle l,E|r,E\rangle|<1$.
If the matrix elements include the internal degrees of freedom such as spin in addition to site indices, we also need the second condition:
\begin{itemize}
\item The pseudospectrum of $H$ under perturbations that only act on internal degrees of freedom stays at the neighborhood of the original spectrum of $H$.
\end{itemize}
(In general, pseudospectrum with a restricted perturbation is called as structured pseudospectrum.) 
Owing to this condition, $|r,E\rangle$ is located far from $|l,E\rangle$ in real space.
To obtain the well-separated zero modes for given $H$, it is efficient to choose the eigenvalue $E$ with the smallest $|\langle l,E|r,E\rangle|$. In the case of the skin modes of the Hatano-Nelson model, all the eigenstates have the same localization length, and the overlap integral $|\langle l,E|r,E\rangle|$ is exponentially small with respect to the size of the system. As a result, the Su-Schrieffer-Heeger model, which is the doubled Hermitian Hamiltonian of the Hatano-Nelson model, has the exponentially localized topological boundary zero modes.

Now, we are in a position to characterize quasi-zero modes of a Hermitian system in terms of nonnormal pseudospectra.
Recall that the boundary modes in semi-infinite dense spectrum of a non-Hermitian system are related to topological boundary zero modes in the corresponding semi-infinite doubled Hermitian Hamiltonian. 
As we mentioned in the previous section, the pseudospectrum looks like the semi-infinite dense spectrum, and Eq. (\ref{semi-pse}) holds in the case of class-A non-Hermitian chains.
In the following, the pseudospectrum plays a similar role as the semi-infinite dense spectrum.
Let us choose $E_{\epsilon}$ in the $\epsilon$ pseudospectrum of $H$:
\begin{align}
    \|(H-E_{\epsilon})|r,E_{\epsilon}\rangle\|&<\epsilon,\notag\\
    \|(H^{\dagger}-E^*_{\epsilon})|l,E_{\epsilon}\rangle\|&<\epsilon.
\end{align}
These inequalities follow from the third definition of the pseudospectrum.
By using these vectors, we obtain
\begin{align}
\left\|
\tilde{H}_{E_{\epsilon}}
\begin{pmatrix}
    0\\
    |r,E_{\epsilon}\rangle
\end{pmatrix}\right\|&<\epsilon,\notag\\
\left\|
\tilde{H}_{E_{\epsilon}}
\begin{pmatrix}
    |l,E_{\epsilon}\rangle\\
    0
\end{pmatrix}\right\|&<\epsilon.\label{pseudozero}
\end{align}
These inequalities indicate that $0$ is in the $\epsilon$ pseudospectrum of the doubled Hamiltonian $\tilde{H}_{E_{\epsilon}}$ from the third definition.
Since the $\epsilon$ pseudospectra of Hermitian matrices are nothing but the $\epsilon$ neighborhood of the spectra, there exist the quasi-zero eigenenergies of $\tilde{H}_{E_{\epsilon}}$ that differ from exact zero by $\mathcal{O}(\epsilon)$, which follows from the second definition. 
In this sense, there is a correspondence between the pseudospectrum of $H$ and the quasi-zero modes of $\tilde{H}_{E_{\epsilon}}$. 
In the case of the Su-Schrieffer-Heeger model, the quasi-zero modes are mainly composed of superposition states of the negative- and positive-chirality modes constructed from the right and left peseudoeigenmodes of the Hatano-Nelson model. In general, pseudoeigenvectors in Eq. (\ref{pseudozero}) are the exact zero modes of some perturbed Hamiltonian $\tilde{H}_{E_{\epsilon}}+\tilde{\eta}$. Conversely, $\tilde{H}_{E_{\epsilon}}$ can be regarded as a Hami
ltonian perturbed from $\tilde{H}_{E_{\epsilon}}+\tilde{\eta}$, which implies that the quasi-zero modes of $\tilde{H}_{E_{\epsilon}}$ are mainly composed of the exact zero modes of $\tilde{H}_{E_{\epsilon}}+\tilde{\eta}$ because of the conventional Hermitian perturbation theory.

Again, these quasi-zero modes are not always anomalous. Whether they are anomalous or not can be roughly decided from the behavior of the pseudospectrum of a nonnormal matrix $H$ (Fig. \ref{fig3}).
Corresponding to a quasi-zero mode associated with $\tilde{H}_{E_\epsilon}$, let us consider $E_\epsilon\in \sigma_\epsilon(H)$. From definition, there exists a matrix $\eta$ with $\|\eta\|<\epsilon $ such that $E_{\epsilon}\in \sigma(H+\eta)$. Thus, $E_{\epsilon}$ is an eigenvalue of $H+\eta$.  Then, if $\sigma_{\epsilon'}(H+\eta)$ with $\epsilon'\gg \epsilon$ is larger than the $\epsilon'$-neighborhood of $\sigma(H+\eta)$, $H+\eta$ still has large nonno
rmality. Thus, $|\langle l,E_{\epsilon}|r,E_{\epsilon}\rangle|$ is less than unity. 
In the case of the Hatano-Nelson model with large but finite system size under the OBC, a pseudoeigenstate with $E_{\epsilon}$ near the OBC eigenspectrum has small localization length, which leads to the small $|\langle l,E_{\epsilon}|r,E_{\epsilon}\rangle|$, while that near the PBC curve is almost delocalized, which leads to $|\langle l,E_{\epsilon}|r,E_{\epsilon}\rangle|\sim1$. These facts are related to the same behavior of boundary modes in the semi-infinite Hatano-Nelson model via Eq. (\ref{semi-pse}). 
Note that these discussions cannot exclude the case where the nonnormal matrix partially behaves like normal one such as the case where the matrix is decomposed into the direct sum of the normal and nonnormal parts.
Thus, $E_{\epsilon}$ should be chosen as the neighborhood of $E$ satisfying $|\langle l,E|r,E\rangle|<1$.

Finally, we discuss the robustness of anomalous nature of zero modes against chiral-symmetry-preserving perturbations.
In the case of the Su-Schrieffer-Heeger model under the OBC, the topological boundary zero modes are fragile against nonlocal perturbations that connect two ends of the chain because of the recombination of the anomalous zero modes into bulk modes, while they are robust against local perturbations described by short-range terms. In our present theory, the generalization of this robustness is given as follows.
\begin{itemize}
\item Anomalous zero modes of $\tilde{H}_E$ are robust against perturbations in the form of 
\begin{align}
    \begin{pmatrix}
    0&\eta\\
    \eta^{\dagger}&0
    \end{pmatrix}
\end{align}
with $\|\eta\|<\epsilon$, if $\sigma(H+\eta)$ stays inside of the $\epsilon$ neighborhood of $\sigma(H)$. 
\end{itemize}
This statement holds because the perturbed $H+\eta$ still has non-normality for the same reason in the previous paragraph. Thus, the anomalous zero modes are robust against such perturbations.
%Note that the general perturbations including nonlocal ones can remove the anomaly if $\epsilon$ is large enough for $\sigma_{\epsilon'}(H+\eta)\sim\sigma_{\epsilon'}(H)$ in the above discussion to be almost the same as the $\epsilon$-neighborhood of the spectrum for any $\epsilon'\gg\epsilon$. The simplest example is again the relationship between the Su-Schrieffer-Heeger model and Hatano-Nelson model.
%In the Su-Schrieffer-Heeger model, the nonlocal perturbations that connect the two ends lead to the recombination of two anomalous zero modes, and they are absorbed into the bulk spectrum. In terms of the Hatano-Nelson model, this phenomenon corresponds to the fact that the nonnormality of the perturbed Hamiltonian $H+\eta$ becomes small for sufficiently large $\epsilon$. 
Roughly speaking, the strength of the anomalous nature of zero modes can be measured by
\begin{align}
    \frac{({\rm Size~ of~ }\epsilon{\rm \mathchar`-pseudospectrum})}{\epsilon}.
\end{align}
The denominator represents the effect of structured perturbations, while the numerator does the effect of general perturbations.
In summary, the nonnormality determines the robustness of the anomalous zero modes.

\subsection{Class-BDI Majorana zero modes from real-structured spectrum and pseudospectrum}
The above discussion can be generalized to other symmetry classes if they support chiral symmetry.
In particular, anomalous zero modes of the superconductors are Majorana fermions, which have lots of applications in condensed matter physics.
We here treat the class-BDI Majorana zero modes as the simplest example.

In the class BDI, both of the time-reversal ($T$) and particle-hole ($C$) symmetries are present:
\begin{align}
    TH^*T^{-1}&=H,~TT^*=1,\\
    CH^*C^{-1}&=-H,~CC^*=1,
\end{align}
where $T$ and $C$ are unitary matrices.
By combining the time-reversal and particle-hole symmetries, we can define the chiral symmetry as in the case of the class AIII.
By taking the Majorana representation $C=1$, where the fermion operators satisfy the Majorana condition $c=c^{\dagger}$, and taking $\Gamma=\sigma_z$, we can rewrite the Hamiltonian as
\begin{align}
    H=
    \begin{pmatrix}
    0&-iR\\
    iR^T&0
    \end{pmatrix},\label{bdimatrix}
\end{align}
where $R$ is a general real square matrix.

The way to generalize the theory developed for the class AIII to the class-BDI Hamiltonians is just imposing the reality on the nonnormal matrices as the additional condition.
Owing to this condition, the pseudospectra in the above discussions should be replaced with the real-structured pseudospectra, where the perturbation matrices are real. To keep the BDI symmetry, the doubled Hamiltonian is constructed by using a real reference point $E$ belonging to the spectra or pseudospectra:
\begin{align}
    \tilde{H}_{E\in\mathbb{R}} := 
\begin{pmatrix}
	0 & -i(R-E) \\
	i(R^{T} - E) & 0 \\ 
\end{pmatrix}.
\end{align}
Note that real eigenvalues do not always exist since complex eigenvalues, accompanied with their conjugates, are allowed to exist.
Thus, the exact zero modes cannot always be defined in this construction, while quasi-zero modes can always be defined for sufficiently large $\epsilon$.

\subsection{Remarks about classes, dimensionality, and gap}
As discussed above, spectra and pseudospectra of nonnormal matrices are related to the Hermitian anomalous zero modes under an additional chiral symmetry. Except for the class-AIII zero modes, the anomalous zero modes in general classes are constructed from the structured nonnormal matrices that obey the symmetry constraints such as the reality in the case of class-BDI zero modes.
Another example is the class DIII, where the corresponding nonnormal spectra and pseudospectra have the Kramers degeneracy owing to the time-reversal symmetry. In terms of the symmetry-protected skin effects, this class has been investigated in Refs. {\cite{Okuma-19,OKSS-20}}.

The dimensionality of the zero modes is also an important factor.
In the case of the topological zero modes in several $D$-dimensional topological insulators/superconductors, symmetry-protected zero modes appear as a part of the $(D-1)$-dimensional gapless Dirac dispersion (or as the defect bound states). In general nonnormal lattice networks, where the topological nature is not always present, such a massless dispersion is possible to be observed as a remnant of higher-dimensional topology. Otherwise, anomalous zero modes originate from the one-dimensional topology embedded in higher-dimensional networks.
The corner states of the higher-order topological phases \cite{BBH-2017} are the typical example of it.

Finally, we note that the present theory does not ensure the presence of the large gap between anomalous zero modes and other states, in contrast to one-dimensional topological insulators and superconductors. In higher-dimensional topological phases with zero modes, there is no gap between the zero modes and nonzero surface states in the thermodynamic limit, while the robustness of zero modes is not affected by such nonzero modes. Similarly, our theory ensures the robustness of anomalous zero modes but does not ensure their isolation from the other states.

\section{Examples of nonnormal network and Hermitian zero modes\label{example}}
\begin{figure*}[]
\begin{center}
　　　\includegraphics[width=17cm,angle=0,clip]{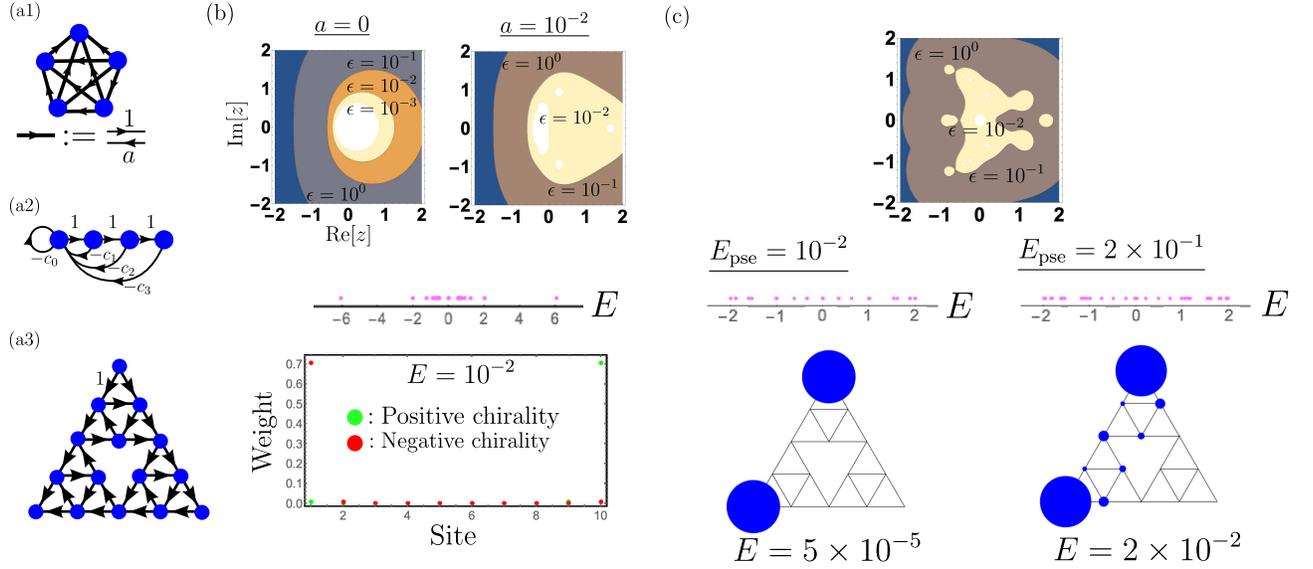}
　　　\caption{(a) Graphical expressions of nonnormal matrices describing (a1) long-range hopping with asymmetric parameter $a$, (a2) algebraic equation $p(x)=0$, and (a3) fractal network. Nodes and directed edges represent the sites and asymmetric hopping terms, respectively.  (b) Pseudospectra of the long-range nonnormal matrix (a1) and corresponding Hermitian spectrum with quasi-zero modes. Contours are defined by $\|[z-H]^{-1}\|=s_{\rm max}([z-H]^{-1})=\epsilon$. The weight function of a quasi-zero mode is separately plotted on positive- and negative-chirality subspaces. (c) Pseudospectrum of the fractal network (a3) for different reference points $E_{\rm pse}$ and corresponding Hermitian spectra with quasi-zero modes. The weight functions of quasi-zero modes are represented by the size of blue spots.}
　　　\label{fig4}
\end{center}
\end{figure*}

One intuitive way to construct Hermitian zero modes is to use the graph-theoretical representation.
In terms of graph theory, the nonnormal matrices can be regarded as adjacency matrices of weighted directed graphs [Fig. \ref{fig4}(a)], where nodes and edges represent the sites including internal degrees of freedom and hopping terms in tight-binding models, respectively. By considering the doubled Hamiltonian, we can construct the class-AIII anomalous zero modes and class-BDI Majorana fermions from complex- and real-weighted directed graphs, respectively.

In this section, we introduce several examples of nonnormal matrices with graphical expressions.
These matrices are useful both for the realization of perturbation-sensitive non-Hermitian systems and construction of Hermitian systems with anomalous zero modes.
We perform the numerical diagonalization for some graphs and compare the results with the pseudospectra obtained by computing the resolvent $\|[z-H]^{-1}\|=s_{\rm max}([z-H]^{-1})$, where $s_{max}(\cdot)$ denotes the largest singular value of the matrix (see the first definition of the pseudospectrum).
In the case of real nonnormal matrices with a real reference point, both of the complex and real (Majorana) fermions can be constructed.
\subsection{Nonnormal networks with all-range hopping}
All-range hopping terms break the short-range nature, which is an implicit but important assumption of the topological bulk-boundary correspondence.
Nevertheless, the anomalous zero modes can still be defined by making use of the nonnormality.
For example, the $L\times L$ nonnormal matrix
\begin{align}
H=
\begin{pmatrix}
0&M&\cdots&M^{L-2}&M^{L-1}\\
aM&0&\cdots&M^{L-3}&M^{L-2}\\
\vdots&\vdots&\ddots&\vdots&\vdots\\
aM^{L-2}&aM^{L-3}&\cdots&0&M\\
aM^{L-1}&aM^{L-2}&\cdots&aM&0
\end{pmatrix},
\end{align}
where $0\leq a<1$, describes the non-Hermitian Hamiltonian with asymmetric all-range hopping terms [Fig. \ref{fig4}(a1)].
The pseudospectra of the system with $L=10, M=1$ for $a=0,10^{-2}$ are plotted in Fig. \ref{fig4}(b), which show the nonnormal behavior [Fig. \ref{fig4}(b)] \footnote{Pseudospectra with $a=0$ and $10^{-2}$ look very different When we regard the case with $a=10^{-2}$ as a perturbed system with respect to that with $a=0$, the norm of this perturbation is estimated as $\epsilon\sim0.06$, which changes the pseudospectrum with $\epsilon$ smaller than this value.}.
For $a=0$, $H$ is defective (non-diagonalizable), and $(1,0,\cdots)^T$ and $(0,\cdots,1)$ are the right and left eigenvectors with eigenvalue $E=0$, respectively.
Thus, the doubled Hamiltonian $\tilde{H}_{E=0}$ has the anomalous exact zero modes completely localized at the boundaries.
For $a\neq0$, on the other hand, $E=0$ is no longer an eigenvalue of $H$, and it is located in the pseudospectrum with $\epsilon=10^{-2}$. Since the pseudospectrum with $\epsilon'=10^{-1}\gg\epsilon$ is much larger than the $\epsilon'$-neighborhood of the $\epsilon$ pseudospectrum, the corresponding pseudoeigenvector still has the large nonnormality (see the previous section for details). Thus, $\tilde{H}_{E=0}$ should have the anomalous quasi-zero modes. In fact, the numerical calculation indicates that there are boundary-localized modes with eigenvalues $\pm10^{-2}$ [Fig. \ref{fig4}(b)].

\subsection{Algebraic equation and nonnormality}
The nonnormality is also related to the algebraic equations.
Solutions of an algebraic equation, or equivalently, roots of a polynomial
\begin{align}
p(x)=:x^{L}+\sum^{L-1}_{i=0}c_ix^i
\end{align}
are sometimes very sensitive to the perturbations including numerical errors.
This sensitivity problem can be understood in terms of spectral theory of a matrix whose characteristic polynomial is $p(x)$:
\begin{align}
H=
\begin{pmatrix}
-c_{L-1}&-c_{L-2}&\cdots&-c_1&-c_0\\
1&0&\cdots&0&0\\
0&1&\cdots&0&0\\
\vdots&\vdots&\ddots&\vdots&\vdots\\
0&0&\cdots&1&0
\end{pmatrix}.
\end{align}
The graphical expression is shown in Fig. \ref{fig4}(a2). 
For example, the companion matrix of the Wilkinson's polynomial $p(x)=\prod^{L}_{i=1}(x-i)$ is a nonnormal matrix with the unstable spectrum and pseudospectrum \cite{Trefethen-97}.

\subsection{Zero modes with power-law decay}
The decay of the localized wave function is not limited to the exponential form.
Let us consider the following square matrix:
\begin{align}
    H=GDG^{-1},
\end{align}
where $G$ is a regular matrix, $D=$diag$(1,2,\cdots,L)$ is a diagonal matrix with natural-number entries, and $L$ is the number of sites. Since this form is nothing but a similarity transformation, the eigenspectrum of $H$ is given by natural numbers. In terms of the nonnormal spectrum, the case with $[G]_{i,j}=(i+j)^{-1}$ has been well studied \cite{Mehrmann-97}. The right eigenvector with the eigenvalue $E_j=j$ is given by
\begin{align}
    (|r,j\rangle)_i=\frac{1}{i+j}.
\end{align}
Thus, the corresponding zero mode with negative chirality of the doubled Hamiltonian is analytically obtained, and it is
localized at $i=1$ with a power-law decay, while the zero mode with positive chirality, on the other hand, has no simple analytical expression.

\subsection{Nonperiodic systems}
While the condensed matter physics focuses mainly on the crystals with translation invariance, several other networks are known to maintain good orders.
Quasicrystals, amorphous, hyperbolic lattices, and fractal structures are typical examples of noncrystalline orders. Here, we consider a nonnormal matrix on a fractal-lattice model (Sierpinski gasket) whose graphical expression is given in Fig. \ref{fig4}(a3). 
The pseudospectrum of this fractal network is much larger than the $\epsilon$ neighborhood of the spectrum [Fig. \ref{fig4}(c)], which indicates strong nonnormality.
By making use of the strong nonnormal behavior of pseudospectrum around the eigenvalue $0$, we construct the quasi-zero modes for different reference points $E_{\rm pse}=10^{-2},2\times10^{-1}$ [Fig. \ref{fig4}(c)]. Owing to the nonnormality, the quasi-zero modes consist of the localized states at different corners.

\section{Summary}
In this paper, we have developed a theory of nonnormality-induced Hermitian zero modes that have a quantum anomaly, including Majorana fermions.
As a generalization of the relationship between topological zero modes and non-Hermitian skin effects, we have considered the doubled Hermitian Hamiltonians constructed from general nonnormal matrices.
We have introduced the spectra and pseudospectra of nonnormal matrices and related them to the anomalous exact and quasi-zero modes of the doubled Hermitian Hamiltonians, which do not have to have topological characterization.
We have shown that the anomalous nature is measured by the amount of nonnormality, and its robustness against the certain perturbations are determined by the behavior of the pseudospectra.
Our theory not only proposes the notion of the nonnormality-protected zero modes, but also adds nonnormal spectra and pseudospectra as tools for topological and non-Hermitian physics.

\acknowledgements
This work was supported by JST CREST Grant No.~JPMJCR19T2, Japan. N.O. was supported by KAKENHI Grant No.~JP18J01610 and JP20K14373 from the JSPS. M.S. was supported by KAKENHI Grant No.~JP20H00131 from the JSPS. 
\appendix

%\bibliographystyle{apsrev4-1}
%\bibliography{NH-topo}

\end{document}